\documentclass [10pt]{article}
\title{Non-time-orthogonality, gravitational orbits, and Thomas precession}
\author{Robert D. Klauber\\1100 University Manor, 38B, Fairfield, Iowa 52556\\email  rklauber@netscape.net}
\date{July 10, 2000}
\usepackage {graphicx}
\usepackage {longtable}

\begin{document}
\maketitle

\bigskip

\begin{abstract}
Non-time-orthogonal analysis of rotating frames is applied to objects in 
gravitational orbits and found to be internally consistent. The object's 
surface speed about its axis of rotation, but not its orbital speed, is 
shown to be readily detectable by any ``enclosed box'' experimenter on the 
surface of such an object. Sagnac type effects manifest readily, but by 
somewhat subtle means. The analysis is extended to objects bound in 
non-gravitational orbit, where it is found to be fully in accord with the 
traditional analysis of Thomas precession.
\end{abstract}

\bigskip

\section*{I. Introduction}

An analysis\footnote{ Robert D. Klauber, ``New perspectives on the 
relatively rotating disk and non-time-orthogonal reference frames'', 
\textit{Found. Phys. Lett. }\textbf{11}(5), 405-443 (1998).} 
$^{,}$\footnote{ Robert D. Klauber, ``Comments regarding recent articles on 
relativistically rotating frames'', \textit{Am. J. Phys.} \textbf{67}(2), 
158-159, (1999).} $^{,}$\footnote{ Robert D. Klauber, ``Non-time-orthogonal 
frames in the theory of relativity'', xxx.lanl.gov paper gr-qc/0005121, 
submitted for publication May 2000.} has been carried out of the 
non-time-orthogonal (NTO) metric obtained when one makes a straightforward 
(and also the most widely accepted) transformation from the lab to a 
relativistically rotating frame. Rather than assuming, as have other 
researchers, that it is then necessary to transform to locally time 
orthogonal frames, one can proceed by considering the NTO metric to be a 
physically valid representation of the rotating frame.

When this is done, one finds time dilation and mass-energy 
dependence\footnote{ Ref 1, pp. 425-429, and ref. 3, eq (5).} on tangential 
speed $\omega $\textit{r} that is identical to the predictions of special 
relativity and the test data from numerous cyclotron experiments. One also 
finds resolutions of paradoxes inherent in the traditional analytical 
treatment of rotating frames. Further, the analysis predicts two different 
experimental results\footnote{ Ref 1, pp. 434-436, and ref 3, section V.B.} 
$^{,}$\footnote{ Ref. 3, Section V.} that, in the context of the traditional 
analysis, have heretofore been considered inexplicable.

One of these results is a persistent non-null signal found by Brillet and 
Hall\footnote{ A. Brillet and J. L. Hall, ``Improved laser test of the 
isotropy of space,'' \textit{Phys. Rev. Lett}., \textbf{42}(9), 549-552 
(1979).} in the most accurate Michelson-Morley type experiment to date, 
which as Aspden\footnote{ H. Aspen, ``Laser interferometry experiments on 
light speed anisotropy,'' \textit{Phys. Lett.,} \textbf{85A}(8,9), 411-414 
(1981).} pointed out, would correspond to an earth surface speed of 
approximately 363 m/sec. It is noteworthy that the earth surface speed at 
the test site is 355 m/sec, and that no other experiment has been sensitive 
enough to test for this effect.

Though the NTO frames analysis predicts such a signal, the question arises 
as to why the earth surface speed should differ from the solar and galactic 
orbital speeds, which yield null signals in the same (and many similar) 
test(s). This article answers this question, as well as a related question 
concerning Thomas precession.

\section*{II. NTO Analysis}

\subsection*{A. Predictions}

NTO frame analysis makes many of the same predictions as the traditional 
analysis for rotating frames, and as emphasized in reference 3, is in accord 
with fundamental principles of relativity theory. Analyses of 
time-orthogonal (TO) frames, including those described by Lorentz, 
Schwarzchild, and Friedman metrics, remains unchanged. The line element 
remains invariant, and differential geometry maintains its reign as 
descriptor of non-inertial systems, whether TO or NTO.

However, NTO analysis does predict some behavior that may seem strange from 
a traditional relativistic standpoint, though it appears corroborated by 
both gedanken and physical experiments\footnote{ Ref 3, section II.} . In 
particular, it was found that velocities in the circumferential direction 
add in a nontraditional way, i.e.

\begin{equation}
\label{eq1}
u_{circum} \,\; = \;\,\,\frac{{ - \omega r + U_{circum} }}{{\sqrt {1 - 
\left( {\omega r} \right)^{2}/c^{2}} }} = \frac{{ - v + U_{circum} }}{{\sqrt 
{1 - v^{2}/c^{2}} }},
\end{equation}

\noindent
where\textit{ u}$_{circum }$is circumferential speed in the rotating frame 
of an object having circumferential speed \textit{U}$_{circum}$ in the 
non-rotating frame, $\omega $ is the angular velocity measured from the 
non-rotating frame, \textit{r} is the radial distance from the center of 
rotation, and \textit{v=}$\omega $\textit{r}. Note that when 
\textit{U}$_{circum}$=0, the object appears in the rotating frame to be 
moving opposite the direction of $\omega $ at speed $\omega $\textit{r} (to 
first order)\textit{,} as is physically reasonable. Time dilation effects 
account for the familiar factor in the denominator.

Expanding on (1), NTO analysis finds the specific result for the speed of 
light in the circumferential direction for rotating (NTO) frames to be 
non-invariant, non-isotropic, and equal to\footnote{ Ref. 1, pg. 425, eq. 
(19) modified by the time dilation factor discussed in the subsequent 
paragraphs to yield physical velocity., and pg. 430, eq. (33).}

\begin{equation}
\label{eq2}
u_{light,circum} \,\; = \;\,\,\frac{{ - \omega r \pm c}}{{\sqrt {1 - \left( 
{\omega r} \right)^{2}/c^{2}} }} = \frac{{ - v \pm c}}{{\sqrt {1 - 
v^{2}/c^{2}} }},
\end{equation}

\noindent
where the sign before \textit{c} depends on the circumferential direction of 
the light ray at \textit{r}. Note the circumferential light speed varies to 
first order with $\omega $\textit{r}.

Relationship (2) leads readily to the prediction of i) the Sagnac\footnote{ 
E.J. Post, "Sagnac effect," \textit{Mod. Phys}. \textbf{39}, 475-493 (1967). 
} effect, and ii) a signal due to the earth surface speed \textit{v} 
precisely like that found by Brillet and Hall. For bodies in gravitational 
orbit (2) does not hold, and $|u_{light,circum} | = c$\textit{,} as in 
traditional relativity. This is because such bodies are in free fall, and 
are essentially inertial, Lorentzian, time orthogonal (TO) frames. They are 
not subject to the idiosyncrasies of non-time-orthogonality, so their 
orbital speed would result in a null Michelson-Morley signal.

\subsection*{B. Orbital Speeds Reconsidered}

While the statements at the end of the preceding subsection may at first 
seem reasonable, under scrutiny they are seen, not only as somewhat 
superficial, but also in apparent conflict with logic used to form the basis 
of the NTO analysis. 

In particular, the gedanken experiment of reference 3 addresses an observer 
fixed to the rim of a rotating disk who sends out two very short pulses of 
light (of length 1/360 of the circumference) that travel in opposite 
directions around the rim of the disk. From the lab frame both light pulses 
have speed \textit{c}. As the two light pulses are traveling the disk is 
rotating ccw, so from the lab frame it is readily apparent that the cw pulse 
strikes the original observer before the ccw pulse. The conclusion reached 
by the disk observer, who knows that both pulses traveled the same distance 
around the rim in his frame, is that in his frame the cw speed of light must 
be greater than the ccw speed of light. This seeming contradiction of the 
relativistic tenet that light speed is invariant, isotropic, and equal to 
\textit{c} is resolved by the NTO analysis leading to (2). That is, in TO 
frames (local, physical) light speed is invariant and isotropic, but in NTO 
frames, such as the rotating frame, it is not.

Applying the same logic to a sun centered rotating frame in which the earth 
is fixed, one would expect the same result, i.e., different ccw and cw light 
speeds as seen from the earth leading to a non-null Michelson-Morley result. 
Yet, as claimed above, the earth in that frame is in gravitational orbit, in 
free fall, and in a Lorentz frame. In such a frame the speed of light must 
always be equal to \textit{c}, and the analysis appears to be internally 
inconsistent.

\section*{III. Resolution of Apparent Inconsistency}

\subsection*{A. Non-spinning Body in Orbit}

As a first step in answering the above conundrum, consider a planet in orbit 
about a star where the planet is not rotating on its own axis relative to 
distant stars, i.e., one solar day equals one year. (See Figure 1.) K is an 
inertial frame with its origin fixed at the center of the sun. The K$_{0}$ 
frame is fixed to the planet and has $\omega $ = 0, though it has orbital 
angular velocity about the sun of $\Omega $ relative to K. \textit{R} is the 
distance in K from the sun center to the planet center, so $\Omega 
$\textit{R }= \textit{V} is the planet's orbital speed in K. For simplicity, 
all velocities are co-planar, the orbit is circular, and unless otherwise 
noted, analysis is confined to first order effects in velocity, time, and 
distance (higher order effects are not measurable.)

\begin{figure}[htbp]
\centering
\includegraphics*[bbllx=0.26in,bblly=0.11in,bburx=6.15in,bbury=3.32in,scale=1.00]{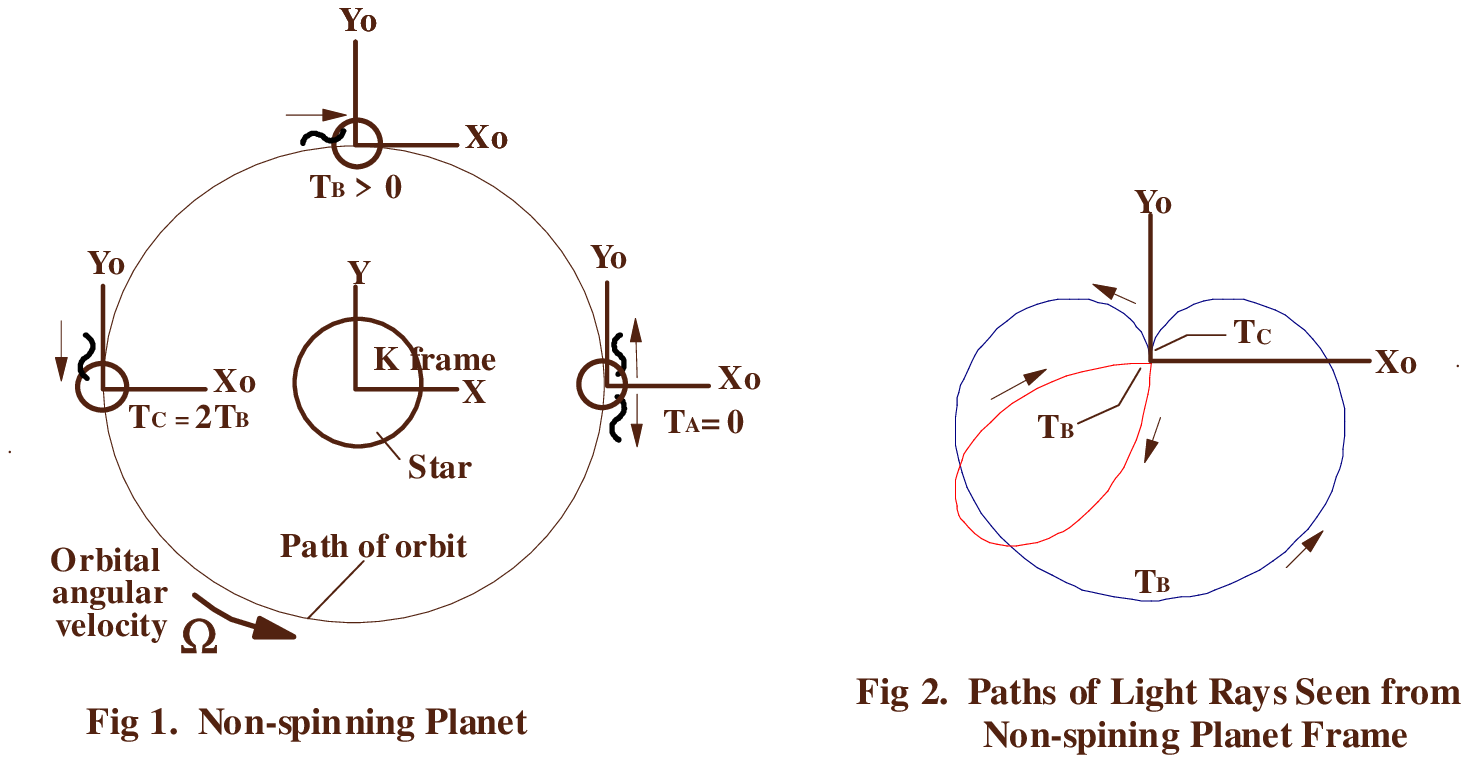}
\end{figure}

\bigskip

Note that an observer on the planet doing experiments (Foucault pendulum, 
Coriolis effects, etc.) inside a closed laboratory (similar to Einstein's 
gedanken intergalactic elevator) would measure zero angular velocity, and be 
unable to determine $\Omega $. 

Now reconsider our gedanken experiment of section II.B with \textit{V} = 
\textit{c}/3. At time T$_{A}$=0, two light rays are emitted from the origin 
of K$_{0}$, one in the "forward" or positive Y$_{0}$ direction and one in 
the "backward" or negative Y$_{0}$ direction. These light rays are reflected 
off of mirrors placed suitably in orbit such that they travel around a 
circumference in K at the orbital radius. In K the speed of light is 
invariant and equal to \textit{c}. Therefore from K we would expect the two 
rays to arrive back at the K$_{0}$ origin at different times, 
\textit{T}$_{B}$ and \textit{T}$_{C}$ (= 2\textit{T}$_{B}$), since K$_{0}$ 
moves along the orbit while the light rays are in transit. 

In K$_{0}$ the speed of light must also be \textit{c}. The conundrum 
dissolves when we note the paths of the two light rays, as depicted in 
Figure 2, are \textit{not} the same as seen from K$_{0.}$ This is because 
K$_{0 }$does not rotate and hence the two light pulses do not travel the 
same circular path as seen from K$_{0}$. In this particular case, the path 
of the ccw pulse is twice that of the cw pulse as seen from 
K$_{0}$\footnote{ This result, as well as Figure 2, can be found by 
inputting the components of the displacement vector from the origin of 
K$_{o}$ to the light rays as a function of time into a spreadsheet or other 
computer program.} . Hence for invariant light speed in K$_{0}$, 
\textit{T}$_{C}$ = 2\textit{T}$_{B}$ as was found in K\footnote{ To simplify 
the discussion we are considering weak gravitational fields, first order 
effects on speed, time, and distance, and hence K$_{0}$ as an effectively 
Lorentzian frame even at distances removed from the K$_{0}$ origin. 
Essentially, we are restricting observable effects to those arising from the 
numerator of (2) and ignoring effects such as that from the denominator, 
which are too subtle to measure in experiments.} .

Note that a Michelson-Morley (MM) interferometer on the surface of the 
planet (i.e., in a Lorentz frame) would detect no variance or anisotropy in 
the speed of light. Yet the Sagnac effect around the orbit would manifest 
completely.

\subsection*{B. Body in Orbit Spinning with $\omega $ = $\Omega $}

Consider next Figure 3 where the planet spins on its own axis (perpendicular 
to the orbital plane) at $\omega $ = $\Omega $. This is similar to the 
rotating disk case in that the same side of the planet faces its sun at all 
times, just as every element on the disk maintains the same alignment 
relative to the disk center. The frame of the rotating planet is designated 
k$_{\omega = \Omega }$.

\begin{figure}[htbp]
\centering
\includegraphics*[bbllx=0.26in,bblly=0.11in,bburx=7.26in,bbury=3.56in,scale=0.90]{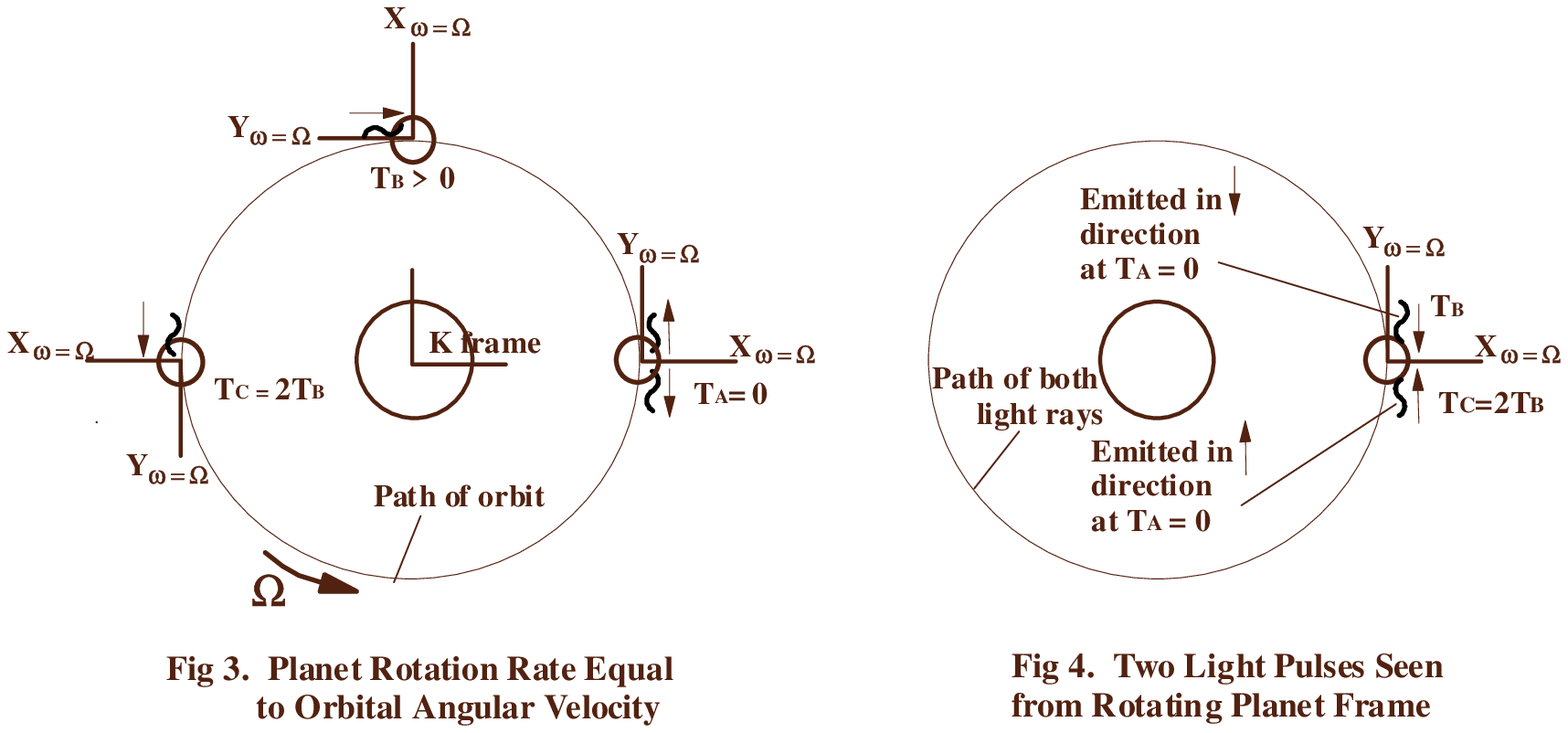}
\end{figure}

\bigskip 

For the frame K, fixed relative to the distant stars, the analysis of two 
light paths in opposite directions around the planet's orbit is identical to 
that for two light rays as seen from the lab frame in the rotating disk case 
of section II.B. For orbital speed of the planet \textit{V} in K equal to 
one-third the speed of light, we would again find 
\textit{T}$_{C}$=2\textit{T}$_{B}$.

From the point of view of the planet frame k$_{\omega = \Omega }$, rotation 
is taking place about the planet's center. Hence the angular velocity our 
experimentalist on the planet would measure would be $\omega   =  
\Omega $. She would then determine the tangent velocity of the planet's 
surface to be

\begin{equation}
\label{eq3}
v_{\omega = \Omega } = \omega r_{p} = \Omega r_{p} 
\end{equation}

\noindent
where \textit{r}$_{p}$ is the radius of the planet, \textit{not }the orbital 
radius about the star. 

Based on NTO analysis, a Michelson-Morley experiment on the surface of the 
planet would find a non-null signal corresponding to the same velocity. 
Hence, a number of independent experiments could detect the planet surface 
tangent velocity, though \textit{not} the orbital tangent velocity.

As seen from the rotating planet frame k$_{\omega = \Omega }$, the two light 
pulses would travel in opposite directions along the same circular path. 
(See Figure 4.) Each pulse has a different speed, according to the +/- sign 
in (2), but with the notable difference that these speeds are now variable 
since \textit{r}, the distance from the planet center to a given light pulse 
is no longer constant. That is, circumferential light speed in the planet 
frame is a function of distance \textit{r} from the origin of the frame to 
the light pulse\textit{, }i.e.,

\begin{equation}
\label{eq4}
u_{light,orbit,k_{\Omega } } \left( {r} \right)\; = \;\,\,\frac{{ - \omega r 
\pm c}}{{\sqrt {1 - \left( {\omega r} \right)^{2}/c^{2}} }} = \frac{{ - 
v\left( {r} \right) \pm c}}{{\sqrt {1 - v\left( {r} \right)^{2}/c^{2}} }}.
\end{equation}

Note that the velocity corresponding to this speed is \textit{perpendicular} 
to the \textbf{r} vector corresponding to distance \textit{r}. That is, (4) 
is only one component of the light velocity along the path traveled as seen 
in k$_{\omega = \Omega }$.

One can then use the two speeds of (4) to determine each pulse speed along 
the paths traveled. When this speed is integrated with respect to time 
between zero and arbitrary time \textit{t}, and the result set equal to the 
distance traveled 2$\pi $\textit{R}, one finds \textit{t }equals (to first 
order) \textit{T}$_{B}$ for the ccw light pulse and \textit{T}$_{C}$ = 
2\textit{T}$_{B}$ for the cw pulse, as before.\footnote{ One does not have 
to actually make this tedious calculation. The times \textit{T}$_{B}$ and 
\textit{T}$_{C}$ have been calculated in the frame K, and those values can 
be transformed to the frame k$_{S }$(introduced in a subsequent paragraph), 
where they will be found to be unchanged to first order. From k$_{S}$ they 
can be again transformed to frame k$_{\omega = \Omega }$, and found 
unchanged once again. The first transformation is shown by both references 1 
and 3 to have \textit{t}$_{s}$=\textit{T}$_{K}$. The second transformation 
is purely a spatial coordinate value shift and has no effect on time.} 

In the k$_{\omega = \Omega }$ frame, the paths traveled are equal, but the 
speeds differ. In the K frame the speeds are equal, but the paths differed. 
In both frames the times of arrival of each pulse are the same.

From the point of view of an observer in a third frame k$_{S}$, centered on 
the sun and rotating at $\Omega ,$ the speeds (again to first order) of 
the two pulses are -$\Omega $\textit{R} +\textit{ c} and -$\Omega $\textit{R 
- c}. (These are not, as mentioned, the speeds a free fall observer in orbit 
at radius \textit{R} would measure.) In k$_{S}$ the path lengths of the two 
light pulses are equal, but the difference in speeds results once again in 
the same values for \textit{T}$_{B}$ and \textit{T}$_{C}$.

\subsection*{C. Comparison with Sagnac Experiment}

The above analysis agrees with the Sagnac experimental results reported by 
Post$^{11}$. In the Sagnac experiment the light beams are not short as in 
our gedanken experiment so arrival times are supplanted by interference 
fringing of the two beams. The number of fringes difference between the two 
beams varies with wave maxima arrival times, and so increases directly with 
the rotational speed. Post\footnote{ Ref. 11, equation (1).} gives the 
empirically determined relative fringe shift $\Delta $\textit{Z=}$\Delta 
$\textit{$\lambda /\lambda $}$_{0}$ as

\begin{equation}
\label{eq5}
\Delta Z = 4\Omega \; \cdot {\bf A}/\lambda _{0} c,
\end{equation}

\noindent
where \textbf{$\Omega $} is the angular velocity, and \textbf{A} is the area 
enclosed by the paths of the two oppositely directly light beams with the 
vector direction orthogonal to the area surface. Although this is an 
empirical result, as noted in reference 1, it can be readily derived from 
(2). 

To find arrival time difference between light beams multiply (5) by 
\textit{$\lambda $}$_{0}$/\textit{c} to yield

\begin{equation}
\label{eq6}
T_{C} - T_{B} = 4\Omega \; \cdot {\bf A}/c^{2}.
\end{equation}

Post acknowledges that (5) (and therefore also (6)) is only accurate to 
first order. He also clearly points out that the area \textit{A} does not 
have to be circular, nor does it have to have the axis of rotation at its 
center. Further, (5) and (6) can be used by observers in either a rotating 
or non-rotating frame, since all quantities used therein are readily 
measured by experimenters in either type of frame. 

Hence, all three frames considered in section III.B would yield the same 
experimental results, since to first order all have the same 
\textbf{\textit{A}} and \textbf{$\Omega $} values. Thus the experimental 
results would be in full accord with the NTO analysis of section III.B.

\subsection*{D. Body in Orbit with Arbitrary $\omega $}

For a situation like the earth where $\omega > > \Omega ,$ or more 
generally for any $\omega $, similar logic would hold. In the former case, 
light following the path of the earth's orbit would seem from the earth 
frame to follow a corkscrew-like path with varying velocity, yet all Sagnac 
type results (interference fringing, arrival times) would be the same 
regardless of the frame from which they are analyzed. The light speed 
anisotropy would in all cases, however, be related to the earth surface 
velocity relative to the inertial frame in which the earth axis is 
stationary. In particular, the earth surface speed at its equator would be 
determined by Michelson-Morley, Foucault pendulum, Coriolis, etc. 
experiments to be

\begin{equation}
\label{eq7}
v_{eq} = \omega r_{eq} ,
\end{equation}

\noindent
where \textit{r}$_{eq}$ is the equatorial radius of the earth.

\section*{IV. Thomas Precession}

\subsection*{A. Traditional Analysis of Thomas Precession}

Figure 1 can be used as an aid in discussing Thomas precession. However, 
instead of a gravitationally bound orbit we now consider Figure 1 to depict 
a charged object such as a classical electron held in orbit by a central 
charge of opposite sign such as an atomic nucleus. Although the orbiting 
K$_{0}$ frame is not spinning, the orbiting (Bohr) electron is spinning 
(unlike the planet in Figure 1) and thereby possesses both intrinsic angular 
momentum and a magnetic moment. Consider that at time \textit{T}=0 a 
projection of the angular momentum vector (which is not necessarily 
orthogonal to the orbital plane) onto the plane of the figure would be 
aligned with the \textit{X}$_{0}$ axis (and hence the \textit{X} axis as 
well).

In Newtonian theory, as the origin of the \textit{X}$_{0}$-\textit{Y}$_{0}$ 
axes orbits the central charge at radius \textit{R}, the angular momentum 
vector orientation remains fixed (relative to K, the frame of the distant 
stars), and so its projection would remain aligned in the direction of the 
\textit{X} axis. In traditional relativity theory, however, the spin 
(angular momentum) axis precesses as seen from K, due to Lorentz contraction 
effects which vary in time due to the oscillating (from the point of view of 
K) centrifugal acceleration. This precession is called Thomas 
precession\footnote{ Edwin F. Taylor and John Archibald Wheeler, 
\textit{Spacetime Physics}, (W.H. Freemand and Co., San Francisco, 1966) pp. 
169-174. Taylor and Wheeler have as lucid and readily assimilable a 
treatment of Thomas precession as any in the literature.} $^{,}$\footnote{ 
George P. Fisher, ``The Thomas Precession'', \textit{Am. J. Phys.}, 
\textbf{40}, 1772-1781 (1972).} $^{,}$\footnote{ A. E. Ruark, and H.C. Urey, 
\textit{Atoms, Molecules, and Quanta }(McGraw-Hill, New York, 1930) pp. 
162-163.} after its discoverer.

Note that the \textit{X}$_{0}$-\textit{Y}$_{0}$ axes frame is not an NTO 
frame, but an accelerating TO frame. It is no different in this regard from 
any frame that undergoes (variable or constant) acceleration without 
rotation. Hence NTO analysis is not in conflict with the traditional 
treatment of Thomas precession for orbital electrons, which is based on TO 
frames only. 

For reference, we note that (to second order), where \textit{v}=$\Omega 
$\textit{R}, Thomas precession equals\footnote{ Ref. 16, equation (134) on 
pg. 173.}

\begin{equation}
\label{eq8}
\omega _{T} \cong - \frac{{1}}{{2}}\frac{{v^{2}}}{{c^{2}}}\Omega ,
\end{equation}

\noindent
and is the precession of the spin vector relative to the \textit{X} axis 
direction of K, as seen from K.

\subsection*{B. Alternative Analysis of Thomas-like Effect}

As is well known, Thomas precession of orbiting electrons alters the 
spin-orbit interaction (which is a function of the precession rate of the 
spin vector) and results in fine structure splitting of atomic spectra. 
There is, however, another way to find the same effect using NTO analysis.

Consider the same spinning electron in the same orbit, but analyze it using 
the orbiting frame k$_{S}$ (which is sun centered and rotating at $\Omega $ 
relative to K). Use the k$_{\omega = \Omega }$ frame (see Figure 3) as a 
convenient local representation of k$_{S}$, and note that the projection of 
the spin angular momentum vector rotates relative to k$_{\omega = \Omega }$ 
(and hence also relative to k$_{S}$) at -$\Omega .  $

According to NTO analysis, there is no Lorentz contraction effect between 
the k$_{S}$ and K frames, and hence no Thomas-like rotation can arise. 
However, there is time dilation in k$_{S}$, and so the rate of rotation of 
the spin vector projection relative to the \textit{X}$_{\omega = \Omega }$ 
axis will not be the same as seen from K as it is seen from k$_{S}$. 
Specifically, if \textit{$\tau $}$_{S}$ is the time on a standard clock in 
k$_{S}$ travelling with the electron for one full rotation of the spin 
vector relative to the \textit{X}$_{\omega = \Omega }$ axis, then the time 
\textit{T}$_{K}$ for the same rotation as seen in K is

\begin{equation}
\label{eq9}
T_{K} = \frac{{\tau _{S} }}{{\sqrt {1 - v^{2}/c^{2}} }}.
\end{equation}

Hence the rotation rates seen in the two frames are related by

\begin{equation}
\label{eq10}
\Omega _{K} = \Omega _{S} \sqrt {1 - v^{2}/c^{2}} \cong \Omega _{S} - 
\frac{{v^{2}}}{{2c^{2}}}\Omega _{S} .
\end{equation}

So

\begin{equation}
\label{eq11}
\Delta \Omega \cong - \frac{{v^{2}}}{{2c^{2}}}\Omega _{S} \cong - 
\frac{{v^{2}}}{{2c^{2}}}\Omega .
\end{equation}

Thus, from (11) one sees the difference in spin vector precession rate 
between that seen in K and that seen in k$_{S}$ found from NTO analysis to 
be the same as that of (8) found from Thomas precession analysis using only 
TO frames. Hence the difference between energy levels measured in the lab 
and the corresponding energy levels that would be measured on a frame 
traveling with the electron is the same for both methods.

\subsection*{C. Rotating Disks and Thomas Precession}

NTO analysis does not, however, predict any type of Thomas precession 
phenomena for objects such as a macroscopic rotating disk in which every 
element of the object rotates with the same angular velocity. According to 
such analyses, no Lorentz contraction effects manifest as internal disk 
stresses. This is in conflict with some analyses based on a more traditional 
approach\footnote{ Daniel P. Whitmire, ``Relativistic Precessions of 
Macroscopic Objects'', \textit{Nature}, \textbf{239}, 207-207 (1972).} , but 
in accord with the Phipps\footnote{ Thomas E. Phipps, Jr., ``Kinematics of a 
Rigid Rotor'', \textit{Nuovo Cimento Lett.}, \textbf{9}, 467-470 (1974).} 
experiment, which found no evidence of the predicted Thomas precession-like 
effects for a spinning disk.

\section*{V. Summary}

We have shown that NTO analysis of bodies in gravitational orbit is 
internally consistent and in agreement with expected results based on the 
Sagnac experiment. The speed of light on bodies in orbit that do not rotate 
relative to distant stars is invariant, isotropic and equal to \textit{c}. 
For a rotating body in orbit, light speed on the surface is anisotropic and 
a function of both the angular velocity of the body and the radial distance 
from the axis of rotation.

NTO analysis does not contravene the traditional analysis of Thomas 
precession for a spinning object held in orbit by a non-gravitational force. 
It does not, however, predict any Thomas precession type effects for an 
object such as a rotating disk wherein every local element in the object 
rotates at the global rotation rate and maintains fixed distance to the axis 
of rotation.

\end{document}